\documentclass{article}
\usepackage{bm}
\usepackage{latexsym}
\usepackage{dcolumn}
\usepackage{amsmath,amsfonts,amssymb}
\usepackage{graphicx,epsfig}
\usepackage{color}

\def\eq#1{{Eq.~(\ref{#1})}}

\newcommand{\mdof}{microscopic degrees of freedom}

\reversemarginpar

\begin{document}

\title{Finite entanglement entropy from the zero-point-area of spacetime}
 \author{T.~Padmanabhan\\
IUCAA, Post Bag 4, Ganeshkhind,\\
 Pune - 411 007, India\\
  \\
email: paddy@iucaa.ernet.in}

\date{ }

\maketitle
\begin{abstract} The calculation of entanglement entropy $S$ of quantum fields in spacetimes with horizon shows that, quite generically, $S$ is (a) proportional to the area $A$ of the horizon and (b) divergent. I argue that this divergence, which arises even in the case of Rindler horizon in flat spacetime, is yet another indication of a deep connection between horizon thermodynamics and gravitational dynamics. In an emergent perspective of gravity, which accommodates this connection, the fluctuations around the equipartition value in the area elements will lead to a minimal quantum of area $\mathcal{O}(1)L_P^2$ which will  act as a regulator for this divergence. In a particular prescription for incorporating the $L_P^2$ as zero-point-area of spacetime, this does happen and the divergence in entanglement entropy is regularized, leading to $S\propto A/L_P^2$ in Einstein gravity. In more general models of gravity, the surface density of microscopic degrees of freedom is different which leads to a  modified regularisation procedure and the possibility that the entanglement entropy --- when appropriately regularised ---  matches  the Wald entropy.

\end{abstract}

\section{Entropy of Horizons versus Temperature of Horizons}

The two key thermodynamic variables that are associated with a black hole horizon are the entropy and temperature. But the manner in which they get associated to a horizon are markedly different and deserves a careful comparison. 

Historically, Bekenstein associated \cite{Bekenstein:1972tm} an entropy with black hole horizon, in order to maintain the validity  of second law of thermodynamics involving the black hole. At that time, the association $S\propto A$ came under criticism because of the prevailing view that the black hole should have zero temperature   to be black. The black hole first acquired the notion of temperature when Hawking's investigation of quantum field theory \cite{Hawking:1975sw} in the black hole spacetime led to a thermal radiation with a temperature $T=1/8\pi M$. In such a calculation the temperature is inferred from the Planck distribution of quanta of the field  but --- given the fact that black hole is radiating these quanta --- it seemed reasonable to attribute this temperature to the black hole. (When we receive photons from the sun, the temperature is a parameter in the Planck distribution of photons but we do attribute this temperature to the solar surface which is radiating the quanta.) One can adopt the valid point of view that the black hole horizon has a temperature $T$ and radiates quanta of all fields at this temperature. 

Similar mathematical procedure allows one to attribute a temperature to \textit{any} horizon near which the metric can be approximated by a Rindler metric. In some cases (like e.g., Rindler  \cite{Davies:1975th} or de Sitter \cite{GHlohia} spacetime) the natural quantum state of the field describes a state of thermal equilibrium rather than a state with a net radiated flux. But there is unanimity of opinion in the literature that all such horizons possess a temperature.

The situation regarding entropy, in contrast, is  unclear. To begin with, one can assign an entropy to the black hole if we \textit{assume} that the result $T=1/8\pi M$ should  hold even if $M$ changes slowly with time and integrate the equation $TdS/dt=dM/dt$. This will lead to the finite result $S=A/4L_P^2$ and there seems to be general agreement that this should be thought of as entropy ``of the black hole" --- though there is no clear idea as to which degrees of freedom of the black hole are involved and where they are located in the spacetime. 

The situation is worse for other horizons. There is no definitive conclusion in the literature as to whether de Sitter horizon or Rindler horizon should have \textit{entropies} associated with them,   in spite of the fact that everybody agrees that \textit{all} horizons have \textit{temperature} $T=\kappa/2\pi$ where $\kappa$ is the surface gravity. (For arguments suggesting that \textit{all} horizons must have entropies associated with them, see e.g.\cite{univent,rop})

There is another crucial difference between the nature of these two thermodynamic variables in the context of horizons. The temperature attributed to the horizon is completely independent of the field equations of the theory. If we have two different models for gravity leading to the same metric (with a horizon) as a solution, the temperature attributed to the horizon will be the same in both models. Temperature is just a property of near horizon geometry and does not know anything about the field equations which the spacetime metric  satisfies.
In contrast, the entropy attributed to the horizon depends on the field equations. This is obvious in the expression for Wald entropy \cite{Wald:1993nt} for a theory based on a general, diffeomorphism invariant, action; but is implicit in any other approach which depends on the physical processes version of first law. Further, the entropy of the horizon is \textit{not} proportional to the area of the horizon in a general theory of gravity. We will come back to implications of this result at the end.

Just as one could attribute the temperature to the quantum field in the presence of a horizon, one can also  assign an  entropy \textit{to the field}. In fact there is a strong argument in favour of this assignment. If we integrate over the field modes on one side of a bifurcation horizon, in the globally defined vacuum state functional of a quantum field, then we get a thermal density matrix $\rho=Z^{-1}\exp (-\beta H)$ with $\beta^{-1}=\kappa/2\pi$ describing the physical processes on the other side \cite{gravitation}. Given the fact that temperature for the quantum field arises from integrating out certain set of modes, it seem reasonable to attribute an entropy to the quantum field due to lack of information about the \textit{same} modes. This is essentially the entanglement entropy of the vacuum state of the field in the presence of a horizon (One could do a similar analysis even in flat spacetime by excising a region of space \cite{comment2};
but the motivation for such a calculation becomes sharper in the presence of a horizon which we will concentrate on.)

 The local redshifted  temperature of the quantum field $T_{\rm loc}$ varies inversely as the proper distance from the horizon $l$ near any horizon which can be approximated by a Rindler metric. Therefore, the entropy density of the thermal quanta varies as $s \propto T_{\rm loc}^3 \propto l^{-3} $ near the horizon in $D=4$. This makes the integrated entropy scale as 
\begin{equation}
 S\propto \int dA_{\perp}\, dl\, l^{-3} \propto \frac{A_{\perp}}{L_c^2}
\label{simple}
\end{equation} 
where $L_c$ is a lower cut-off length. We see that the result is proportional to the area of the horizon but quadratically divergent. This analysis depends only on the validity of the Rindler approximation near the horizon and is independent of the field equations of the theory  which the metric might satisfy. 

More formally, the entanglement entropy is given by $S=-{\rm Tr} (\hat \rho \ln \hat\rho)$ where $\hat\rho=\rho/Z$ is the normalized density matrix with $Z={\rm Tr} \rho$ being the partition function.  This can be calculated using the alternative form:
\begin{equation}
S=-(\alpha\partial_\alpha-1)\ln {\rm Tr}  \rho^\alpha\vert_{\alpha=1}
\end{equation} 
The ${\rm Tr} \rho^\alpha$ can be determined using the `replica trick' \cite{replica} and can be related to the effective action (or free energy of the theory) which in turn can be expressed in terms of the Schwinger proper time Kernel $K(x,y;s)$ (`heat Kernel') of the theory \cite{kernel}. For a free, massless, scalar field in $D$ dimensional Euclidean space, this leads to the expression:
\begin{equation}
 S=\frac{A_{D-2}}{12}\int_0^\infty \frac{ds}{s}K_{D-2}(x,x;s)
\label{SfromK}
\end{equation} 
where $A_{D-2}$ is the transverse area  (see e.g., \cite{solo}). The coincidence limit of 
the Kernel  behaves as $K_{D-2}(x,x;s)\propto s^{-(D-2)/2}$ and hence the integral in \eq{SfromK} diverges as $L_{c}^{-(D-2)}$ at the lower limit where $L_{c}$ is a lower cutoff length scale. In $D=4$ this gives $S\propto A_\perp/L_{c}^2$, which diverges quadratically as in \eq{simple}. Extensive studies of entanglement entropy have shown that the above two features are very robust: (a) The leading term in $S$ is proportional to the area of the horizon
 and (b) $S$ is divergent; in 4 dimensions it is quadratically divergent. 

At this stage one usually introduces a lower limit cut-off $L_c\approx L_P$ at Planck length and obtains $S\propto A_\perp/L_P^2$ in $D=4.$ While most people seem to believe that $L_P$ should provide a regulator to the entanglement entropy, such a prescription  has far reaching implications which I will now elaborate upon. 

\section{Entanglement Entropy and Microstructure of Spacetime}

To see this, note that, even in  the absence of gravity ($G_N=L_P^2=0$), one can study quantum field theory in an inertial and Rindler frame and obtain the result that the horizon is endowed with a temperature. In the conventional perspective this result has nothing to do with gravity and $G$ never appears in the result. 
If we now compute the entanglement entropy, it will turn out to be divergent even in the simple context of a free field theory
which is exactly what \eq{simple} or \eq{SfromK} tell us. 

\textit{Since the free field theory in flat spacetime knows nothing about gravity or entropy of black holes, how would we handle this divergent result?}

In particular, in the absence of gravity (and Planck length) how would we regularize the entanglement entropy? 
This difficulty can be tackled at a fundamental level \textit{only if} there  exists a deeper connection between the Rindler horizon thermodynamics and the microscopic structure of spacetime which supplies the quantum of area $L_P^2$. That is, `free' field theories in Rindler spacetime must know about the existence of gravity arising from Planck scale spacetime microstructure
(It is sometimes argued \cite{comment3} that the tracing of \textit{all} the modes on one side of the horizon has no operational significance and this is why $S$ is divergent. Even then, one needs to (indirectly) invoke gravitational effects to limit the operational significance of measurements, without which there is no way of getting $L_P$ in to the analysis.)  

In fact, considerable amount of evidence has accumulated  over the years suggesting such a connection between horizon thermodynamics and microstructure of spacetime and indicating that gravity is better described as an emergent phenomenon like elasticity or fluid mechanics \cite{rop,others}.  In particular, it has been shown that: (a) The  field equations of gravity reduce to a thermodynamic identity on the horizons in a wide variety of models much more general than just Einstein's gravity \cite{ronggen}.
(b) It is possible to obtain \cite{aseemtp} the field equations of gravity --- again for a wide class of theories --- from purely thermodynamic considerations by extremising a suitable entropy density for spacetime.
  
  In this paradigm, one considers spacetime (described by the metric, curvature etc.) as a physical system analogous to a gas or a fluid (described by density, velocity etc.). The fact that either physical system (spacetime or gas) exhibits thermal phenomena shows  that there \textit{must} exist microstructure in either system. Therefore one does not try to quantize gravity but instead attempts to provide a quantum description of spacetime. This is identical in spirit to the fact that one does not quantize, say, the variables in the Navier-Stokes equation (which is analogous to the gravitational field equation) to obtain a quantum theory of matter but instead identifies the appropriate microscopic degrees of freedom (molecules, atoms, ....) and develops a quantum theory of \textit{these} degrees of freedom. 
We do not yet know what are the correct microscopic degrees of freedom of the spacetime; but the horizon thermodynamics provides a clue along the following lines.

This connection between macroscopic thermodynamics and the existence of microscopic degrees of freedom comes out  clearly --- for both gas and spacetime --- in the equipartition law $\Delta E = (1/2) (\Delta n) k_B T$ connecting the number of degrees of freedom $\Delta n$ required to store and energy $\Delta E$ at the temperature $T$.  In the case of a gas, $\Delta n$ scales as the volume of the substance and essentially counts the number of molecules. The finiteness of $\Delta n=\Delta E/(1/2)k_BT$ shows the breakdown of continuum description and is a direct proof of discrete microstructure in the gaseous system. \textit{It has been shown recently \cite{avogadro} that an identical relation holds for the spacetime in a wide class of gravitational theories.} In the case of Einstein gravity in $D=4$ the result can be expressed in the form:
 \begin{equation}
\Delta E=\frac{1}{2}
 (\Delta n) k_BT;\quad \Delta n \equiv  \frac{\sqrt{\sigma}\,d^2 x}{ L_P^2}=\frac{\Delta A}{L_P^2}
 \label{idn}
\end{equation}
where $\Delta A = \sqrt{\sigma} d^2 x$ is patch of proper area of a two-surface. So, in the context of Einstein's theory, we find that the microscopic degrees of freedom $\Delta n$ scales in proportion with area --- unlike  gaseous systems in which $\Delta n$ will scale as volume. (This is closely related to the `holographic' nature of gravitational action principles \cite{holo}). 
This result shows that 
the number of \mdof\ in an element of area $ A $ is $ A / L_P^2$ which is exactly what one would have expected if 
there is a quantum of area  $L_P^2$.  The  fluctuations in the \mdof\  will now lead to a dispersion $\delta A$ in the area with the bound $\delta A> \mathcal{O}(1)L_P^2$. 

\section{Zero-Point Area as a regulator for entanglement entropy}

In a more complete description one would expect these fluctuations to be incorporated into the Kernel $K_{D-2}(x,x,s)$ in \eq{SfromK} so that $L_P^2$ arises as a natural cut-off and makes the entanglement entropy finite.  Given the structure of \eq{SfromK}, the answer will depend on the conjectured modification of the theory at Planck energies and --- in fact --- the regularization is not \cite{solo} assured for all possible modifications. We shall consider a specific prescription of regularizing the theory 
and show that it \textit{does} lead to finite entanglement entropy. 

This prescription is based on the conjecture that quantum gravitational fluctuations can be incorporated into the theory by making the path integral `duality invariant' between a path of length $l$ and one of length $L^2/l$ where $L=\mathcal{O}(1) L_P$.  This involves replacing $l$ by $[l+(L^2/l)]$ in the relativistic path integrals. One can show \cite{first} that this is equivalent to modifying the standard Schwinger Kernel as follows:
\begin{equation}
K(x,y;s)\rightarrow
 K(x,y;s)\exp(-L^2/s)
\label{prescription}                       
\end{equation} 
which introduces an exponentially strong regularization near $s=0$ in the integrals involving the Kernel.

This prescription was  suggested in ref.\cite{first} and its consequences  (including the connection with string theory) were explored in several subsequent papers \cite{follow} which describe
 the motivation and justification for this prescription in detail. I will not repeat them here   except to recall three features which is relevant for our discussion. 

(a) Let $l^2(x,y|g_{ab})$ be the square of the proper length between two events $x,y$ (along some curve) in a spacetime with metric $g_{ab}$. If the metric undergoes quantum fluctuations, around a mean value $\bar g_{ab}$ one can define a mean value $\langle l^2(x,y)\rangle $ by averaging over the metric fluctuations. We will then expect \cite{lpbound}:
\begin{equation}
\langle l^2(x,y)\rangle \approx \bar l^2(x,y)+\mathcal{O}(1)L_P^2 
\label{minL} 
\end{equation} 
in the limit of 
 $x\to y$,
where the first term is the classical, mean, value due to the metric $\bar g_{ab}$ and the second term is the dispersion around this value due to fluctuations which gives a `zero-point-area' $L_P^2$. It can be shown that \cite{first} the prescription of path integral duality is equivalent to the introduction of such a zero-point-area to the spacetime. This matches with the area fluctuations arising from equipartition if we interpret the second term in \eq{minL} as the minimal  fluctuations in the \mdof .

(b) When we consider quantum gravitational fluctuations around the flat spacetime, this effect should make the coincidence limit of Green functions finite. This is precisely what happens with the prescription that modifies $K(x,y;s)$ to
 $K(x,y;s)\exp(-L^2/s)$. The Euclidean Green function now gets modified as:
\begin{eqnarray}
\label{modgreen}
G(x,y) &\propto& \int_0^\infty ds \ K(x,y,s)\propto\frac{1}{(x-y)^2}\\
&\to& \int_0^\infty ds \ K(x,y,s) e^{-L^2/s}\propto\frac{1}{(x-y)^2+4L^2}\nonumber
\end{eqnarray}
for a massless field.  The finiteness of the coincidence limit of $G(x,x)$ is a non-perturbative result and cannot be obtained by a Taylor series expansion in $(x-y)^2/L^2$.

(c) To avoid misunderstanding, it should be stressed that \eq{prescription} is a prescription to incorporate quantum structure of the spacetime and cannot be derived from a local, unitary, Lorentz invariant, field theory. In particular, it is not a heat Kernel of a quantum field theory with a suitably modified Green function. For example, one can easily evaluate (see \cite{first}) the Fourier transform $G(p^2)\equiv 1/F(p^2)$ of the modified Green function in \eq{modgreen} --- which can be expressed in terms of Bessel functions ---and construct a field theory based on the operator $F(\square)$. Such a modified field theory will have a heat Kernel $\mathcal{G}(x-y;s)\equiv\langle x|\exp -sF(\square)|y \rangle$. This heat kernel, however, will \textit{not} be the same as the one obtained by the prescription in \eq{prescription}. This is obvious from the fact that Fourier transform $\mathcal{G}(p,s)$ of $\mathcal{G}(x-y,s)$ in $(x-y)$ has the form $\exp[-sF(p^2)]$ instead of the standard form for free massless field, $\exp(-sp^2)$. But in our prescription, $K(p,s)\propto\exp[-sp^2-(L^2/s)]$, which, of course, cannot be expressed in the form $\exp[-sF(p^2)]$.
While the prescription in in \eq{prescription} modifies the Green function, it is not true that the modified Green function can be used to reconstruct the prescription in \eq{prescription} in terms of a modified field theory.

We can now compute the entanglement entropy with 
our prescription using the 
 modified Kernel  $K(x,y;s)\exp(-L^2/s)$ in place of $K(x,y;s)$ in \eq{SfromK}. The integrals are trivial and we get a finite result:
\begin{equation}
 S=\frac{1}{12}\left( \frac{1}{4\pi}\right)^{(D-2)/2}\left( \frac{A}{L^{D-2}}\right)
\end{equation} 
which reduces, in $D=4$ to
\begin{equation}
S=\frac{A}{48\pi L^2}=\frac{A}{4L_P^2} 
\end{equation} 
if we set $12\pi L^2=L_P^2$. Of course, without a more fundamental theory we cannot determine $L$ independently; but  we can now determine the cut-off parameter in path integral duality prescription if we demand $S=(1/4)(A_\perp/L_P^2)$. The key point is that the result is finite, unlike in some other modifications of the high energy sector, based on modified \textit{field thoeries}, considered for example in ref. \cite{solo}. (This paper considers modifications in which the Fourier transform $K(p,s)$ of $K(x,s)$ in $x$ has the form $\exp[-sF(p^2)]$ instead of the standard form $\exp(-sp^2)$. But, as we said earlier, in our prescription, $K(p,s)\propto\exp[-sp^2-(L^2/s)]$, which, of course, cannot be expressed in the form $\exp[-sF(p^2)]$. The fact that even drastically modifying the field theory --- by using an operator $F(\square)$) instead of $\square$ --- does not lead to finite entanglement entropy \textit{strengthens our conjecture that the solution to this infinity needs to be found at a deeper level}.) The same calculation can also be performed for the BTZ black hole in (1+2) dimensions using the same prescription  and one obtains a similar, finite, result \cite{dklsk}. In a complete description, $G_N$ will get renormalized and this has also been computed with the above prescription (see the first paper in \cite{follow}). The scaling due to number of species of fields can be incorporated into this correction. None of these affects our conclusions.

The conceptual structure which  now emerges has the following ingredients: (i) The entanglement entropy is divergent even in flat, Rindler spacetime QFT  in the absence of gravity. (ii) Its regularization demands the existence of  a deeper connection between horizon thermodynamics and gravity, which is present in the emergent paradigm of gravity. (iii) In this approach, one can determine   the the surface density of spacetime degrees of freedom and show that it obeys the equipartition law $\Delta E = (1/2) (\Delta n) k_B T$. (iv) The  fluctuations in these degrees of freedom around equipartition value will lead to a
zero-point-area in spacetime in \eq{minL}, which  can be incorporated into the field theory by a suitable modification of the Kernel.
(v) This, in turn, regularizes the entanglement entropy, closing the logical loop.

\section{Further Generalisations: Can entanglement entropy match Wald entropy in general?}

There have been several speculations in the literature as to  whether the entanglement entropy itself can account for the entropy of the horizon. The key difficulty with such an identification, in the \textit{conventional} perspective, is the following: 
Given a metric which has a Rindler approximation near the horizon, the leading order term in entanglement entropy will be proportional to $A_{\perp}$ (once some kind of regularization is introduced). But as I mentioned earlier, the entropy of the horizon depends on the field equations of the theory   which the metric satisfies and is, in general, given by the Wald entropy \cite{Wald:1993nt}. It is unlikely that the QFT of matter in a given metric will have sufficient information to produce an entanglement entropy which will identically match with the Wald entropy. So, unless we believe gravity must be described by Einstein's theory, we cannot identify entanglement entropy of matter fields with horizon entropy.

It may be possible that the regularization procedure (which is always needed) might also lead to equality of Wald entropy and entanglement entropy.  This is because the regularization prescription itself should depend on the theory of gravity. For example, one motivation for $L_P^2$ acting as a regulator in \eq{minL} comes from the fact that there is an operational limitation \cite{lpbound} to measuring shorter length scales in Einstein gravity if we demand that the energy $E=c\hbar/L$ involved in probing a length $L$ should satisfy the black hole radius bound $GE/c^4<L$ (This lack of precision in the location of a boundary may be required, in any case, to have finite entanglement entropy in QFT under certain circumstances; see e.g.,\cite{cassini}). Such a bound will change in other models of gravity since e.g., the black hole radius of energy $E$ will change. 

One can address this modification in the emergent paradigm of gravity, which generalizes in a very natural manner to more general theories of gravity. The surface density of microscopic degrees of freedom in these theories is given \cite{avogadro} by a relation similar to \eq{idn} with 
\begin{equation}
\Delta n=32\pi P^{ab}_{cd}\epsilon_{ab}\epsilon^{cd} \Delta A                                                             \end{equation} 
where $\epsilon_{ab}$ is the binormal in the transverse case and 
$
P^{abcd} \equiv (\partial L/\partial R_{abcd}).
$
It can be shown  that this counting of microscopic degrees of freedom leads precisely to the Wald entropy of the horizon in these models \cite{avogadro}. But the microscopic fluctuations around equipartition value are also now different and --- if the correct model of gravity is different from Einstein's theory ---  we need to modify \eq{minL} correspondingly. In Einstein gravity, $32\pi P^{ab}_{cd}\epsilon_{ab}\epsilon^{cd}=L_P^{-2}$ and in more general theories, 
\begin{equation}
L_{eff}^{-2}\equiv 32\pi P^{ab}_{cd}\epsilon_{ab}\epsilon^{cd}                                                                \end{equation} 
will replace $L_P^{-2}$. The entanglement entropy $\Delta S\propto \Delta A/L_{eff}^2$, regularised with $L_{eff}^{-2}$ will match with Wald entropy of a patch of horizon.   

One simple example is the $f(R)$ theories of gravity with $f(0)=0$ which also has Schwarzschild metric as a solution but with an effective gravitational constant scaled by $G_N^{-1}\to f'(0)G_N^{-1}$ so that the effective Planck length also gets renormalized to $L_{eff}^{-2}=f'(0)L_P^{-2}$. The Wald entropy will  now be $S_{Wald}=(1/4)(f'(0)A/L_P^2)$. (There is even a claim  that Wald entropy is always $(1/4)$th of area when measured in units of effective coupling constant $G_{eff}$; see \cite{always4}).
 If we now regularize the divergence in the entanglement entropy with the renormalized Planck length, then the Wald and entanglement entropies will match.  
To implement this idea rigorously, we need  a regularization prescription \textit{for the Kernel} obtained by extending the ideas of \cite{first} to a general theory of gravity.  This question is under investigation.

I thank H. Cassini,  T. Jacobson, D. Kothawala, S. Shankaranarayanan and L. Sriramkumar for comments on the manuscript.

\end{document}